\documentclass[10pt,amsmath,amssymb,pra,superscriptaddress,twocolumn]{revtex4}
\usepackage{graphicx}
\usepackage{dcolumn}
\usepackage{bm}
\usepackage{times}
\usepackage{subfigure}
\usepackage{booktabs}
\usepackage{color}
\newcounter{one}
\usepackage{amssymb,amsmath,theorem} 
\newtheorem{theorem}{Theorem}
\newtheorem{lemma}{Lemma}

\newcommand{\affA}{
Department of Physics, The University of Tokyo, Komaba, Meguro, Tokyo 153-8505 
}

\begin{document}
\title{Second law of the information thermodynamics with entanglement transfer}
\author{Hiroyasu Tajima}
\affiliation{\affA}

\begin{abstract}
We present a new inequality which holds in the thermodynamical processes with measurement and feedback controls with using only the Helmholtz free energy and the entanglement of formation:
$W_{\mathrm{ext}}\le-\Delta F-k_{B}T\Delta E_{F}$.
The quantity $-\Delta E_{F}$, which is positive, expresses the amount of entanglement transfer from the system $S$ to the probe $P$ through the interaction $\hat{U}_{SP}$ during the measurement.
It is easier to achieve the upper bound in the new inequality than in the Sagawa-Ueda inequality \cite{sagawa1}.
The new inequality has clear physical meaning: in the above thermodynamical processes, the work which we can extract from the thermodynamic system is greater than the upper bound in the conventional thermodynamics by the amount of the entanglement extracted by the measurement.
\end{abstract}
 
\maketitle

\section{Introduction}
The second law of thermodynamics appears to be violated in thermodynamic processes that include measurements and feedbacks. This well-known fact has been the center of attention and numerous studies have long been conducted on such processes \cite{oldresult1,oldresult2,oldresult3,oldresult4,oldresult5,sagawa1,jacobs,sagawa2,hatten1,hatten2,hatten3,hatten4}.
The second law of information thermodynamics \cite{sagawa1} derived by Sagawa and Ueda is a monumental landmark of such studies;
in the case of an isothermal process, it is expressed as
\begin{equation}
W_{\mathrm{ext}}\le -\Delta F +k_{B}TI_{\mathrm{QC}},\label{sagawainequality}
\end{equation}
where $I_{\mathrm{QC}}$ is the QC-mutual information content \cite{sagawa1}. 
This inequality gives a new upper bound for the work extracted from a thermodynamic system when  measurement and feedback are permitted on the system.

When the measurement is classical ($\left[\hat{\rho},\sqrt{\hat{M}^{\dag}_{(k)}\hat{M}_{(k)}}\right]=0$, where $\{\hat{M}_{(k)}\}$ is the measurement and $\hat{\rho}$ is the density matrix of the system with the baths), the QC-mutual information reduces to the classical mutual information content.
Therefore, in the classical world, the work that we can extract from information thermodynamic processes is greater than the upper bound of the conventional thermodynamics by the amount of information which we obtain from the measurement. 
On the other hand, when the measurement is not classical, the physical meaning of the QC-mutual information is unclear.
We will also show below that when we use finite systems for the heat baths, the upper bound of \eqref{sagawainequality} is not necessarily achievable.

In this paper, we present a new information thermodynamic inequality with using only the Helmholtz free energy and the entanglement of formation:
\begin{equation}
W_{\mathrm{ext}}\le -\Delta F -k_{B}T\Delta E_{F},\label{preinequality}
\end{equation}
where the difference $\Delta E_{F}$ of the entanglement is taken between before and after the unitary interaction $\hat{U}_{SP}$ between the system $S$ and the probe $P$ during the measurement.
The quantity $-\Delta E_{F}$ is always nonnegative and expresses the amount of entanglement transfer from $S$ to $P$ through $\hat{U}_{SP}$. 
Hence, the inequality \eqref{preinequality} has clear physical meaning: the work that we can extract from information thermodynamic processes is greater than the upper bound of the conventional thermodynamics by the amount of entanglement which we obtain from the measurement.
In other words, from a thermodynamical point of view, we can interpret the entanglement transfer as the information transfer.
In the above context, we introduce a new information content $I_{\mathrm{E}}=-\Delta E_{F}$, to which we refer as the entanglement information.
It has a clear physical meaning even when $I_{\mathrm{QC}}$ does not.
We also show that, the condition for the achievement of the upper bound of the new
inequality \eqref{preinequality} is looser than that of the inequality \eqref{sagawainequality}. 

\section{set up of the whole system}
As the setup, we consider a thermodynamic system $S$ that is in contact with heat baths $B_{m}$ for $m=1,2,...,n$ which are at temperatures $T_{1}$,…,$T_{n}$, respectively.
We refer to the whole set of heat baths $\{B_{m}\}$ as $B$.
Except when we perform measurement or feedback control, we express the Hamiltonian of the whole system as 
\begin{equation}
\hat{H}(\vec{\lambda}(t))=\hat{H}^S(\vec{\lambda}^{S}(t))+\sum^{n}_{m=1}[\hat{H}^{SB_{m}}(\vec{\lambda}^{SB_{m}}(t))+\hat{H}^{B_{m}}],
\end{equation}
where $\hat{H}^{S}(\vec{\lambda}^{S}(t))$ is the Hamiltonian of the system $S$, $\hat{H}^{B_{m}}$ is the Hamiltonian of the bath $B_{m}$, and $\hat{H}^{SB_{m}}(\vec{\lambda}^{SB_{m}}(t))$ is the interaction Hamiltonian between the system $S$ and the heat bath $B_{m}$.
The Hamiltonian is controlled through the external parameters $\vec{\lambda}^{S}(t)$ and $\vec{\lambda}^{SB_{m}}(t)$.
We assume that there exists a value of $\vec{\lambda}^{SB_{m}}(t)=\vec{\lambda}_{0}$ such that $\hat{H}^{SB_{m}}(\vec{\lambda}_{0})=\hat{0}$.
We call the time evolution of the whole system with controlled values of $\vec{\lambda}^{S}(t)$ and $\vec{\lambda}^{SB_{m}}(t)$ a thermodynamic operation.
We further assume that we can realize a thermodynamic equilibrium state at temperature $T_{m}$ by connecting $S$ and $B_{m}$ and waiting.
Note that the equilibrium state may not be a canonical distribution.
We define the energy $U$ of a state $\hat{\rho}$ as $\mathrm{tr}[\hat{\rho}\hat{H}]$ and define the Helmholtz free energy $F$ for an equilibrium state at a temperature $T$ as $-k_{B}T\log Z(\beta)$, where $\beta\equiv(k_{B}T)^{-1}$ and $Z(\beta)\equiv \mathrm{tr}[\mathrm{exp}(-\beta\hat{H})]$.

\section{information thermodynamic process}

\begin{figure}
\includegraphics[clip, scale=0.5]{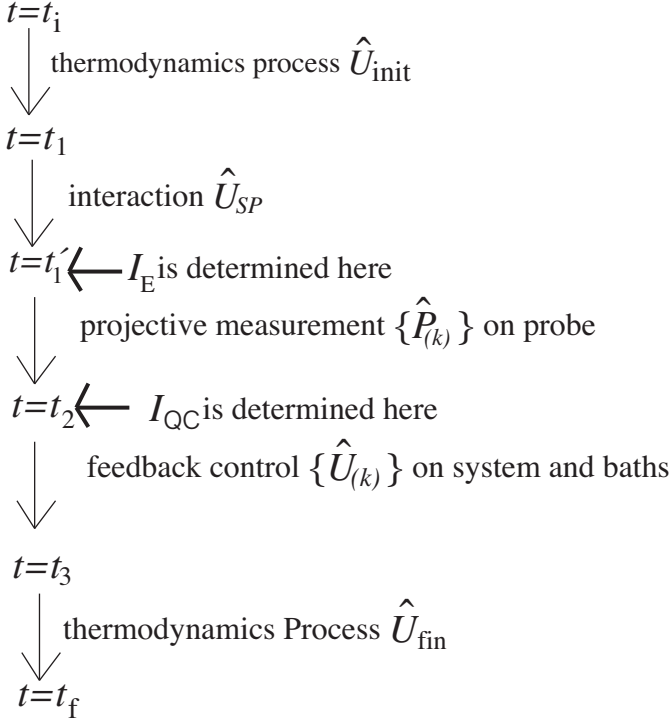}
\caption{Schematic of the thermodynamic processes from $t_{\mathrm{i}}$ to $t_{\mathrm{f}}$.}
\label{process}
\end{figure}

Under the setup in Section 2, we consider the following thermodynamic processes from $t=t_{\mathrm{i}}$ to $t=t_{\mathrm{f}}$ (Fig. \ref{process}).
At $t=t_{\mathrm{i}}$, we start the process with the following canonical initial state:
\begin{eqnarray}
\hat{\rho}_{\mathrm{i}}=\frac{\mathrm{exp}[-\beta\hat{H}^{S}_{\mathrm{i}}]}{Z^{S}_{\mathrm{i}}(\beta)}\otimes\frac{\mathrm{exp}[-\beta_{1}\hat{H}^{B_{1}}]}{Z^{B_{1}}(\beta_{1})}\otimes...\otimes\frac{\mathrm{exp}[-\beta_{n}\hat{H}^{B_{n}}]}{Z^{B_{n}}(\beta_{n})},
\end{eqnarray}
where $\hat{H}^{S}_{\mathrm{i}}=\hat{H}^S(\vec{\lambda}^{S}(t_{\mathrm{i}}))$, $Z^{S}_{\mathrm{i}}=\mathrm{tr}[\mathrm{exp}[-\beta\hat{H}^{S}_{\mathrm{i}}]]$, $\beta_{m}=(k_{B}T_{m})^{-1}$ and $Z^{{B}_{m}}=\mathrm{tr}[\mathrm{exp}[-\beta_{m}\hat{H}^{B_{m}}]]$ $(m=1,...,n)$.
From $t=t_{\mathrm{i}}$ to $t=t_{1}$, we perform a thermodynamic operation $\hat{U}_{\mathrm{init}}$.
At $t=t_{1}$, the state is therefore given by $\hat{\rho}_{1}=\hat{U}_{\mathrm{init}}\hat{\rho}_{\mathrm{i}}\hat{U}^\dag_{\mathrm{init}}$.
Adding a proper reference system $R$, we can find a pure state $\left|\psi_{SBR}\right\rangle$ which satisfies 
\begin{equation}
\mathrm{tr}_{R}[\left|\psi_{SBR}\right\rangle\left\langle\psi_{SBR}\right|]=\hat{\rho}_{1}.
\end{equation}
From $t=t_{1}$ to $t=t'_{1}$, we introduce a unitary interaction $\hat{U}_{SP}$ between the system $S$ and the probe $P$, which is initialized to a state $\left|0_{P}\right\rangle$.
At $t=t'_{1}$, the state of the whole system is expressed as
\begin{equation}
\left|\psi_{PSBR}\right\rangle=(\hat{U}_{SP}\otimes\hat{1}_{B}\otimes\hat{1}_{R})(\left|0_{P}\right\rangle\otimes\left|\psi_{SBR}\right\rangle),\label{PSBR}
\end{equation}
where $\hat{1}_{B}$ and $\hat{1}_{R}$ are the identity operators.

At this point, we define the new quantity $I_{\mathrm{E}}$, namely the entanglement information,  as follows:
\begin{eqnarray}
I_{\mathrm{E}}&\equiv& E^{SB\mbox{-}R}_{F}(\left|\psi_{SBR}\right\rangle\left\langle\psi_{SBR}\right|)-E^{SB\mbox{-}R}_{F}(\hat{\rho}_{SBR})\label{I_{E}}\\
&=&S(\hat{\rho}_{1})-E^{SB\mbox{-}R}_{F}(\hat{\rho}_{SBR}),\nonumber
\end{eqnarray}
where 
\begin{equation}
S(\hat{\rho})\equiv-\mathrm{tr}[\hat{\rho}\log \hat{\rho}], \enskip \hat{\rho}_{SBR}\equiv \mathrm{tr}_{P}[\left|\psi_{PSBR}\right\rangle\left\langle\psi_{PSBR}\right|],
\end{equation}
and $E^{SB\mbox{-}R}_{F}(\hat{\rho})$ is the entanglement of formation \cite{E-formation} between $SB$ and $R$;
\begin{equation}
E^{SB\mbox{-}R}_{F}(\hat{\rho})\equiv\min_{\hat{\rho}_{SBR}=\sum q_{j}\left|\phi^{j}\right\rangle\left\langle\phi^{j}\right|} \sum_{j}q_{j}E^{SB\mbox{-}R}(\left|\phi^{j}\right\rangle)
\end{equation}
with $E^{SB\mbox{-}R}(\left|\phi^{j}\right\rangle)$ being the entanglement entropy \cite{E-entropy} between $SB$ and $R$ for a pure state $\left|\phi^{j}\right\rangle$.
Note that $E^{SB\mbox{-}R}_{F}(\left|\psi_{SBR}\right\rangle\left\langle\psi_{SBR}\right|)$ and $E^{SB\mbox{-}R}_{F}(\hat{\rho}_{SBR})$ indicate the amount of entanglement between $SB$ and $R$ at $t=t_{1}$ and $t=t'_{1}$, respectively.
To put it simply, we can express $I_{\mathrm{E}}$ as follows:  
\begin{eqnarray}
I_{\mathrm{E}}&\equiv& -\Delta E^{SB\mbox{-}R}_{F}\nonumber\\
&=&E^{SB\mbox{-}R}_{F}|_{\textrm{before}\enskip\hat{U}_{SP}}-E^{SB\mbox{-}R}_{F}|_{\textrm{after}\enskip\hat{U}_{SP}}.\label{defIE}
\end{eqnarray}
We also note that $E^{SB\mbox{-}R}_{F}(\left|\psi_{SBR}\right\rangle\left\langle\psi_{SBR}\right|)$ is equal to the amount of entanglement between $SB$ and the rest of the whole system at $t=t'_{1}$.
Thus, we can interpret $I_{\mathrm{E}}$ as the amount of entanglement between $SB$ and $R$ that is taken by the probe $P$ during the interaction $\hat{U}_{SP}$.

From $t=t'_{1}$ to $t=t_{2}$, we perform a projective measurement $\{\hat{P}_{(k)}=\sum_{i}\left|k,i_{P}\right\rangle\left\langle k,i_{P}\right|\}$ on the probe, where $\{\left|k,i_{P}\right\rangle\}$ are pure states of the probe. 
At $t=t_{2}$, we obtain a result $k$ with probability $p_{k}$, and then the state of $SB$ becomes
\begin{equation}
\hat{\rho}_{2}=\sum_{k}p_{k}\hat{\rho}^{(k)}_{2}=\sum_{i,k}q_{i,k}\hat{\rho}^{ik}_{2},
\end{equation}
where 
\begin{eqnarray}
\sqrt{q_{i,k}}\left|\psi^{ik}_{SBR}\right\rangle&=&(\left\langle k,i_{P}\right|\otimes \hat{1}_{SBR})\left|\psi_{PSBR}\right\rangle,\\
\hat{\rho}^{ik}_{2}&=&\mathrm{tr}_{R}[\left|\psi^{ik}_{SBR}\right\rangle\left\langle\psi^{ik}_{SBR}\right|],\\
p_{k}\hat{\rho}^{(k)}_{2}&=&\mathrm{tr}_{PR}[\sum_{i}q_{i,k}\left|\psi^{ik}_{SBR}\right\rangle\left\langle\psi^{ik}_{SBR}\right|],
\end{eqnarray}
with $\left|\psi^{ik}_{SBR}\right\rangle$ and $\hat{\rho}^{(k)}_{2}$ being normalized.
We can interpret the above as performing a measurement $\{\hat{M}_{(k)}\}$, where
\begin{equation}
\hat{M}_{(k)}=\sum_{i}\left\langle k,i_{P}\right|\hat{U}_{SP}\left|0_{P}\right\rangle,
\end{equation}
on $S$ from $t=t_{1}$ to $t=t_{2}$.
The QC-mutual information \cite{sagawa1,sagawa2} is determined here for the first time. It is  expressed as
\begin{eqnarray}
I_{\mathrm{QC}}&\equiv& S(\hat{\rho}_{1})+H\{p_{k}\}\nonumber\\
&&+\sum_{k}\mbox{tr}[\sqrt{\hat{D}_{k}}\hat{\rho}_{SB}\sqrt{\hat{D}_{k}}\log \sqrt{\hat{D}_{k}}\hat{\rho}_{SB}\sqrt{\hat{D}_{k}}],\label{I_{QC}}
\end{eqnarray}
where $\hat{D}_{k}\equiv \hat{M}^\dag_{(k)}\hat{M}_{(k)}$.

We emphasize the following two points.
First, we can determine the unitary interaction $\hat{U}_{SP}$ and the projective measurements $\hat{P}_{(k)}$ for any measurement $\hat{M}_{(k)}$.
Hence, if we can evaluate the QC-mutual information $I_{\mathrm{QC}}$, then we can also evaluate $I_{\mathrm{E}}$. 
Second, the timings at which $I_{\mathrm{E}}$ and $I_{\mathrm{QC}}$ are defined are different.
The information $I_{\mathrm{E}}$ is defined when only $\hat{U}_{SP}$ is completed, whereas the information $I_{\mathrm{QC}}$ is defined when the measurement $\{\hat{M}_{(k)}\}$ is also completed.
Thus, for two measurements with the same $\hat{U}_{SP}$, $I_{\mathrm{E}}$ takes the same value but $I_{\mathrm{QC}}$ may take different values.

From $t=t_{2}$ to $t=t_{3}$, we perform a feedback control depending on the measurement result $k$.
To be precise, we perform a unitary transformations $\hat{U}_{(k)}$ on $SB$.
At $t=t_{3}$, the state of $SB$ is given by
\begin{equation}
\hat{\rho}_{3}\equiv \sum_{k}p_{k}\hat{U}_{(k)}\hat{\rho}^{(k)}_{2}\hat{U}^\dag_{(k)}.\label{rho3}
\end{equation} 
From $t=t_{3}$ to $t=t_{\mathrm{f}}$, we choose a thermodynamic operation $\hat{U}_{\mathrm{fin}}$ whose final state is assumed to be equilibrium and perform it.
We also assume that by $t_{\mathrm{f}}$ system S and heat bath $B_{m}$ will have reached thermodynamic equilibrium at temperatures $T'$ and $T_{m}$, respectively.
Note that we only assume that the final state is macroscopically equilibrium; the final state may not be a canonical distribution given by
\begin{equation}
\hat{\rho}^{\mathrm{can}}_{\mathrm{f}}\equiv\frac{\mathrm{exp}[-\beta'\hat{H}^{S}_{\mathrm{f}}]}{Z^{S}_{\mathrm{f}}(\beta')}\otimes\frac{\mathrm{exp}[-\beta_{1}\hat{H}^{B_{1}}]}{Z^{B_{1}}(\beta_{1})}\otimes...\otimes\frac{\mathrm{exp}[-\beta_{n}\hat{H}^{B_{n}}]}{Z^{B_{n}}(\beta_{n})},\label{rhofcan}
\end{equation}
where $\beta'$ is the inverse temperature of the final state of the system.
We hereafter call the above process as the information thermodynamic process.

\section{main results}
For the above information thermodynamic processes, we present five results.
The first theorem is the new second law of information thermodynamics:

\begin{theorem}
For any information thermodynamic process, the following inequality holds:
\begin{equation}
\frac{U^{S}-F^{S}}{T}+\sum^{n}_{m=1}\frac{Q_{m}}{T_{m}}\le -\frac{U'^{S}-F'^{S}}{T'}+k_{B}I_{\mathrm{E}},\label{new2ndlaw}
\end{equation}
where $Q_{m}\equiv \mathrm{tr}[\hat{H}^{B_{m}}(\rho_{\mathrm{i}}-\rho_{\mathrm{f}})]$ and 
 the quantities $U^{S}$, $F^{S}$, $U'^{S}$ and $F'^{S}$ the energy and the Helmholtz free energy at $t_{\mathrm{i}}$ and $t_{\mathrm{f}}$, respectively. 
When the the system undergoes an isothermal process in contact with a single heat bath $B$ at temperature $T$, the inequality \eqref{new2ndlaw} reduces to 
\begin{eqnarray}
W_{\mathrm{ext}}&\le& -\Delta F +k_{B}TI_{\mathrm{E}}\label{tajimainequality}\\
&=&-\Delta F -k_{B}T\Delta E^{SB-R}_{F}
\end{eqnarray}
\end{theorem}

The second theorem shows that we can always achieve the upper bound of \eqref{new2ndlaw} when we use infinite systems for the heat baths.

\begin{theorem}
When we use infinite systems as the heat baths, there is at least one set of projective measurement $\{\hat{P}_{(k)}\}$ and feedback $\{\hat{U}_{(k)}\}$ which achieve the upper bound of \eqref{new2ndlaw} for any interaction $\hat{U}_{SP}$.
\end{theorem}

With the third, fourth and fifth results, we will see that the condition for the achievement of the upper bound of the new inequality \eqref{new2ndlaw} is looser than that of the inequality
\begin{equation}
\frac{U^{S}-F^{S}}{T}+\sum^{n}_{m=1}\frac{Q_{m}}{T_{m}}\le -\frac{U'^{S}-F'^{S}}{T'}+k_{B}I_{\mathrm{QC}},\label{old2ndlaw}
\end{equation}
 when we use finite systems for the heat baths. 
First, let us present the third result:

\begin{theorem}
If we can always achieve the upper bound of \eqref{old2ndlaw} with a proper feedback $\{\hat{U}_{(k)}\}$, we can always achieve the upper bound of \eqref{new2ndlaw} with a proper set of projective measurement $\{\hat{P}_{(k)}\}$ and feedback $\{\hat{U}_{(k)}\}$.
\end{theorem}

The fourth and fifth results show that the converse of Theorem 3 is \textit{not} true.

\begin{theorem}
When the following conditions are satisfied, we can always achieve the equality of \eqref{new2ndlaw} with proper choices of $\{\hat{P}_{(k)}\}$ and $\{\hat{U}_{(k)}\}$ for any $\hat{U}_{SP}$:

\textit{Condition} $1$: The system $S$ is a two-level system.

\textit{Condition} $2$: The thermodynamic operations $\hat{U}_{\mathrm{init}}$ and $\hat{U}_{\mathrm{fin}}$ satisfy
the equation of the following inequality
\begin{equation}
\frac{U^{S}-F^{S}}{T}+\sum^{n}_{m=1}\frac{Q_{m}}{T_{m}}\le -\frac{U'^{S}-F'^{S}}{T'}.\label{s-q}
\end{equation}

\textit{Condition} $3$: $\vec{\lambda}^{SB}(t)=\vec{\lambda}_{0}$ is satisfied for $t_{1}\le t\le t_{2}$.
\end{theorem}
Condition 2 dictates that we do not waste energy during the thermodynamic processes.
Condition 3 implies that the system and baths do not interact during the measurement; if the system and baths interact during the measurement, the information obtained by the probe contains the information about the system as well as about the baths. 
Thus, we can interpret Theorem 4 as follows:
We can completely use the information obtained by the probe with a proper interpretation $\{\hat{P}_{(k)}\}$ and a proper feedback $\{\hat{U}_{(k)}\}$, if we do not waste energy during the thermodynamic processes and if the information describes only the system.

\begin{theorem}
Under Conditions 1--3, there is a measurement $\{\hat{M}_{(k)}\}$ for which we cannot achieve the equality of \eqref{old2ndlaw}
with any $\hat{U}_{(k)}$.
\end{theorem}

\section{proofs of main results}
Let us prove Theorems $1$--$5$.

\textbf{Proof of Theorem 1}:
Theorem 1 is directly given by the following lemma:
\begin{lemma}
For any measurement $\{\hat{M}_{(k)}\}$, the following inequality holds:
\begin{equation}
I_{\mathrm{QC}}\le I_{\mathrm{E}}.\label{IQCleIE}
\end{equation}
\end{lemma} 
The inequalities \eqref{old2ndlaw} and \eqref{IQCleIE} give \eqref{new2ndlaw}.
Lemma $1$ and Theorems $4$ and $5$ seem to contradict each other.
Though the upper bound of \eqref{new2ndlaw} is always achievable and though the inequality \eqref{IQCleIE} exists, there is a case in which the upper bound of \eqref{old2ndlaw} is not achievable. 
However, the contradiction is only spurious.
Note that when $I_{\mathrm{E}}$ is determined, we can take $\{\hat{P}_{(k)}\}$ freely;
in other words, we can choose the ``best" interpretation of the information obtained by the probe.
On the other hand, when $I_{\mathrm{QC}}$ is determined, $\{\hat{P}_{(k)}\}$ is also determined already, and thus our interpretation of the probe's information is fixed uniquely.

Let us prove Lemma 1.
Because of the definitions \eqref{I_{E}} and \eqref{I_{QC}}, we prove
\begin{eqnarray}
E^{SB\mbox{-}R}_{F}(\hat{\rho}_{SBR})&\le& -H\{p_{k}\}\\
&-&\sum_{k}\mbox{tr}[\sqrt{\hat{D}_{k}}\hat{\rho}_{SB}\sqrt{\hat{D}_{k}}\log \sqrt{\hat{D}_{k}}\hat{\rho}_{SB}\sqrt{\hat{D}_{k}}].\nonumber
\end{eqnarray}
We can express the above as follows:
\begin{eqnarray}
&&-H\{p_{k}\}-\sum_{k}\mbox{tr}[\sqrt{\hat{D}_{k}}\hat{\rho}_{SB}\sqrt{\hat{D}_{k}}\log \sqrt{\hat{D}_{k}}\hat{\rho}_{SB}\sqrt{\hat{D}_{k}}]\nonumber\\
&=&\sum_{k}p_{k}S(\hat{\rho}^{(k)}_{2})=\sum_{k}p_{k}S(\sum_{i}\frac{q_{i,k}}{p_{k}}\hat{\rho}^{(ik)}_{2})\nonumber\\
&\ge& \sum_{i,k}q_{i,k}S(\hat{\rho}^{(ik)}_{2}) = \sum_{i,k} q_{i,k} E^{SB\mbox{-}R}(\left|\psi^{ik}_{SBR}\right\rangle) \nonumber\\
&\ge& E^{SB\mbox{-}R}_{F}(\hat{\rho}_{SBR}),
\end{eqnarray}
where 
\begin{equation}
\hat{\rho}_{SBR}=\sum_{i,k}q_{i,k}\left|\psi^{ik}_{SBR}\right\rangle\left\langle\psi^{ik}_{SBR}\right|=\mathrm{tr}_{P}[\left|\psi_{PSBR}\right\rangle\left\langle\psi_{PSBR}\right|].
\end{equation}
$\Box$

\textbf{Proof of Theorems 2 and 3}:
First we prove Theorem 3.
Let us take an ensemble $\{q_{(k)},\left|\psi^{(k)}_{SBR}\right\rangle\}$ which satisfies
\begin{eqnarray}
E^{SB\mbox{-}R}_{F}(\hat{\rho}_{SBR})&\equiv&\sum_{k}q_{k}E^{SB\mbox{-}R}(\left|\phi^{k}_{SBR}\right\rangle),\\
\hat{\rho}_{SBR}&=&\sum q_{k}\left|\phi^{k}_{SBR}\right\rangle\left\langle\phi^{k}_{SBR}\right|.
\end{eqnarray}
Then, we can take an orthonormal basis $\{\left|k_{P}\right\rangle\}$ which satisfies
\begin{equation}
\left|\psi_{PSBR}\right\rangle=\sum_{k} \sqrt{q_{k}}\left|k_{P}\right\rangle\left|\phi^{k}_{SBR}\right\rangle.
\end{equation}
Let us take the projective measurement $\{\hat{P}_{(k)}\}$ as $\{\left|k_{P}\right\rangle\left\langle k_{P}\right|\}$.
Then, $p_{k}$ reduces to $q_{k}$, and thus
\begin{eqnarray}
I_{\mathrm{QC}}&\equiv& S(\hat{\rho}_{1})+H\{p_{k}\}\nonumber\\
&&+\sum_{k}\mbox{tr}[\sqrt{\hat{D}_{k}}\hat{\rho}_{SB}\sqrt{\hat{D}_{k}}\log \sqrt{\hat{D}_{k}}\hat{\rho}_{SB}\sqrt{\hat{D}_{k}}],\nonumber\\
&=& S(\hat{\rho}_{1})-\sum_{k}p_{k}S(\hat{\rho}^{(k)}_{2})\nonumber\\
&=& S(\hat{\rho}_{1})-\sum_{k}q_{k}E^{SB\mbox{-}R}(\left|\phi^{k}_{SBR}\right\rangle)\nonumber\\
&=& S(\hat{\rho}_{1})-E^{SB\mbox{-}R}_{F}(\hat{\rho}_{SBR})=I_{\mathrm{E}}.
\end{eqnarray}
Thus, for an arbitrary unitary $\hat{U}_{SP}$, there is a projective measurement $\{\hat{P}_{(k)}\}$ that satisfies 
$I_{\mathrm{E}}=I_{\mathrm{QC}}$.

Theorem 2 directly follows from Theorem 3.
When we use infinite systems for the heat baths, we can always achieve the upper bound of \eqref{old2ndlaw} for any measurement $\{\hat{M}_{(k)}\}$ \cite{jacobs}.
Because of the above and Theorem 3, we can always achieve the upper bound of \eqref{new2ndlaw}. $\Box$

\textbf{Proof of Theorem 4}:
As in the derivation of \eqref{old2ndlaw} in Ref. \cite{sagawa1}, we can obtain the inequality \eqref{new2ndlaw} by transforming 
\begin{equation}
S(\hat{\rho}_{\mathrm{i}})\le -\mathrm{tr}[\hat{\rho}_{\mathrm{f}}\log \hat{\rho}^{\mathrm{can}}_{\mathrm{f}}]+I_{\mathrm{E}}.\label{targetresult2}
\end{equation}
Thus, we only have to prove that for any $\hat{U}_{SP}$, we can always take $\{\hat{P}_{(k)}\}$ and $\{\hat{U}_{(k)}\}$ that satisfy
\begin{equation}
S(\hat{\rho}_{\mathrm{i}})= -\mathrm{tr}[\hat{\rho}_{\mathrm{f}}\log \hat{\rho}^{\mathrm{can}}_{\mathrm{f}}]+I_{\mathrm{E}}.\label{targetresult2'}
\end{equation}

First, we prove that if $\hat{\rho}_{3}$ in \eqref{rho3} is a canonical distribution, we can transform \eqref{targetresult2'} into
\begin{equation}
E^{S\mbox{-}R}_{F}(\hat{\rho}_{SR})=S(\hat{\rho}^{S}_{3}),\label{targetresult2fin}
\end{equation}
where $\hat{\rho}_{SR}\equiv \mathrm{tr}_{B}[\hat{\rho}_{SBR}]$, $\hat{\rho}^S_{3}\equiv \mathrm{tr}_{B}[\hat{\rho}_{3}]$, and $E^{S\mbox{-}R}_{F}$ is the entanglement of formation between $S$ and $R$.
Thanks to \eqref{I_{E}} and $S(\hat{\rho}_{\mathrm{i}})=S(\hat{\rho}_{1})$, we can transform \eqref{targetresult2'} into
\begin{equation}
0= -\mathrm{tr}[\hat{\rho}_{\mathrm{f}}\log \hat{\rho}^{\mathrm{can}}_{\mathrm{f}}]-E^{SB\mbox{-}R}_{F}(\hat{\rho}_{SBR}).\label{targetresult2''}
\end{equation} 
Note that a thermodynamic operation from a canonical distribution to an equilibrium state achieves the equality of \eqref{s-q} if and only if the final state is a canonical distribution too [6].
Thus, because of Condition 2, if $\hat{\rho}_{3}$ is a canonical distribution, $\hat{\rho}_{\mathrm{f}}$ is the canonical distribution $\hat{\rho}^{\mathrm{can}}_{\mathrm{f}}$ in \eqref{rhofcan}. Then we can transform \eqref{targetresult2''} into
\begin{equation}
E^{SB\mbox{-}R}_{F}(\hat{\rho}_{SBR})= S(\hat{\rho}_{3}),\label{targetresult2'''}
\end{equation}
where we use $S(\hat{\rho}_{3})=S(\hat{\rho}_{\mathrm{f}})$.
Thus, we only have to transform \eqref{targetresult2'''} into \eqref{targetresult2fin}.
If the following three equations hold, \eqref{targetresult2'''} and \eqref{targetresult2fin} are equivalent;
\begin{eqnarray}
E^{SB\mbox{-}R}_{F}(\hat{\rho}_{SBR})&=&E^{S\mbox{-}R_{1}}_{F}(\hat{\rho}_{SR_{1}})+E^{B\mbox{-}R_{2}}(\left|\psi_{BR_{2}}\right\rangle),\label{targetresult2''''}\\
E^{S\mbox{-}R}_{F}(\hat{\rho}_{SR})&=&E^{S\mbox{-}R_{1}}_{F}(\hat{\rho}_{SR_{1}})\label{targetresult2omake},\\
S(\hat{\rho}_{3})&=&S(\hat{\rho}^{S}_{3})+E^{B\mbox{-}R_{2}}(\left|\psi_{BR_{2}}\right\rangle),\label{targetresult2'''''}
\end{eqnarray}
where we divide $R$ into a two-level subsystem $R_{1}$ and the rest $R_{2}$. 

Let us first prove \eqref{targetresult2''''}.
Owing to Condition 2 and the fact that $\hat{\rho}_{\mathrm{i}}$ is a canonical distribution, $\hat{\rho}_{1}$ is a canonical distribution as well.
Because of Condition 3, $\hat{H}(t_{1})=\hat{H}^S\otimes\hat{H}^B$ is valid.
Thus, because of Condition 1, under the proper basis of $R$ we can divide $R$ into a two-level subsystem $R_{1}$ and a subsystem $R_{2}$ and express $\left|\psi_{SBR}\right\rangle$ as follows:
\begin{equation}
\left|\psi_{SBR}\right\rangle=\left|\psi_{SR_{1}}\right\rangle\otimes\left|\psi_{BR_{2}}\right\rangle.\label{productpre}
\end{equation}
Owing to \eqref{PSBR} and \eqref{productpre}, we can express $\left|\psi_{PSBR}\right\rangle$ as follows:
$\left|\psi_{PSBR}\right\rangle=\left|\psi_{PSR_{1}}\right\rangle\otimes\left|\psi_{BR_{2}}\right\rangle$.
Thus, we can express $\hat{\rho}_{SBR}$ as
\begin{equation}
\hat{\rho}_{SBR}=\hat{\rho}_{SR_{1}}\otimes\left|\psi_{BR_{2}}\right\rangle\left\langle\psi_{BR_{2}}\right|,\label{product2}
\end{equation}
where $\hat{\rho}_{SR_{1}}\equiv \mathrm{tr}_{P}[\left|\psi_{PSR_{1}}\right\rangle\left\langle\psi_{PSR_{1}}\right|]$.
We have \eqref{targetresult2''''} and \eqref{targetresult2omake} from \eqref{product2}.

Next, we prove \eqref{targetresult2'''''}.
Note that $S$ has been isolated for $t_{1}\le t\le t_{3}$ with proper $\{\hat{U}_{(k)}\}$.
We can therefore express $\hat{\rho}_{3}$ as $\hat{\rho}^{S}_{3}\otimes\left|\psi'_{BR_{2}}\right\rangle\left\langle\psi'_{BR_{2}}\right|$ with $E^{B\mbox{-}R_{2}}(\left|\psi_{BR_{2}}\right\rangle)=E^{B\mbox{-}R_{2}}(\left|\psi'_{BR_{2}}\right\rangle)$.
Thus, we have \eqref{targetresult2'''''}.

Now, we only have to find $\{\hat{P}_{(k)}\}$ on $P$ and $\{\hat{U}_{(k)}\}$ on $S$ such that $\hat{\rho}_{3}$ and $\hat{\rho}^{S}_{3}$ are canonical distributions and that \eqref{targetresult2fin} holds.
We first prove that if $\hat{\rho}^{S}_{3}$ is a canonical distribution, $\hat{\rho}_{3}$ is also a canonical distribution.
To prove this, we only have to note that $\mathrm{tr}_{R_{2}}[\left|\psi'_{BR_{2}}\right\rangle\left\langle\psi'_{BR_{2}}\right|]$ is a canonical distribution because $B$ has been isolated for $t_{1}\le t\le t_{3}$ and because $\mathrm{tr}_{S}[\hat{\rho}_{1}]$ is a canonical distribution.
We second find $\{\hat{P}_{(k)}\}$ on $P$ and $\{\hat{U}_{(k)}\}$ on $S$ such that  $\hat{\rho}^{S}_{3}$ is a canonical distribution and that \eqref{targetresult2fin} holds.
Because both $S$ and $R_{1}$ are two-level systems, we can treat the state $\left|\psi_{PSR_{1}}\right\rangle$ as a three-qubit pure state under a proper basis of $P$.
In Appendix, we prove the following with the approach used in Ref. \cite{tajima}:
we can perform a projective measurement $\{\tilde{P}_{(k)}\}_{k=0,1}$ on the probe $P$ such that the results $\tilde{P}_{(k)}\left|\psi_{PSR_{1}}\right\rangle$ are LU-equivalent for $k=0,1$ and 
$E^{S\mbox{-}R_{1}}(\tilde{P}_{(k)}\left|\psi_{PSR_{1}}\right\rangle)=E^{S\mbox{-}R_{1}}_{F}(\hat{\rho}_{SR_{1}})$
is valid.
Because the results $\tilde{P}_{(k)}\left|\psi_{PSR_{1}}\right\rangle$ are LU-equivalent, there exists $\{\hat{V}_{(k)}\}_{k=0,1}$ on $S$, which satisfies 
$\hat{\rho}^{\mathrm{pre}}_{3}\equiv \hat{V}_{(k)}\hat{\rho}^{S(k)}_{2}\hat{V}^\dag_{(k)}$,
where $\hat{\rho}^{S(k)}_{2}\equiv \mathrm{tr}_{R_{1}}[\tilde{P}_{(k)}\left|\psi_{PSR_{1}}\right\rangle\left\langle\psi_{PSR_{1}}\right|\tilde{P}_{(k)}]$.
Owing to Condition 1, if $E^{S\mbox{-}R_{1}}(\tilde{P}_{(k)}\left|\psi_{PSR_{1}}\right\rangle)\ne0$, we can make the state $\hat{\rho}^{S}_{3}=\hat{V}\hat{\rho}^{\mathrm{pre}}_{3}\hat{V}^\dag$  a canonical distribution with a unitary transformation $\hat{V}$ on $S$.
Thus, $\{\tilde{P}_{(k)}\}$ and $\{\hat{V}_{(k)}\}$ are the measurement and feedback that we want.$\Box$

\textbf{Proof of Theorem 5}:
It is sufficient to prove the existence of a counterexample of the measurement $\{\hat{M}_{(k)}\}$.
The equality of \eqref{old2ndlaw} is valid only if there exists a set of unitary transformations $\{\hat{U}_{(k)}\}$ that satisfy $\sum_{k}p_{k}S(\hat{\rho}^{(k)}_{2})=S(\hat{\rho}_{3})$ \cite{sagawa1}.
We can transform $S(\hat{\rho}_{3})$ as follows:
\begin{eqnarray}
S(\hat{\rho}_{3})&=&S(\sum_{k=0,1}p_{k}\hat{U}_{(k)}\hat{\rho}^{(k)}_{2}\hat{U}^\dag_{(k)})\label{henkeiresult3}\\
&=&\sum_{k=0,1}p_{k}S(\hat{\rho}^{(k)}_{2})+\sum_{k=0,1}p_{k}D(\hat{U}_{(k)}\hat{\rho}^{(k)}_{2}\hat{U}^\dag_{(k)}||\hat{\rho}_{3}),\nonumber
\end{eqnarray}
where $D(\hat{\rho}||\hat{\rho}')\equiv \mathrm{tr}[\hat{\rho}(\log \hat{\rho}-\log \hat{\rho}')]$.
Because $D(\hat{\rho}||\hat{\rho}')=0$ if and only if $\hat{\rho}=\hat{\rho}'$ and because of \eqref{henkeiresult3}, the equation $\sum_{k}p_{k}S(\hat{\rho}^{(k)}_{2})=S(\hat{\rho}_{3})$ is valid if and only if $\{\hat{\rho}^{(k)}_{2}\}$ are LU-equivalent for $k=0,1$, in other words, if and only if the measurement $\{\hat{M}_{(k)}\}$ is a deterministic measurement.
Because of this logic, if Theorem 5 were not valid, any measurement $\{\hat{M}_{(k)}\}$ would be deterministic.
This is clearly false, and thus Theorem 5 holds.$\Box$

\section{conclusion}
To conclude, we obtain a new information thermodynamic inequality.
In this inequality, the information gain is the entanglement gain;
the new information content $I_{\mathrm{E}}$ represents the amount of the entanglement between the system and the reference system which the probe takes from the system.
The new information content depends only on the premeasurement state of the system and the unitary interaction between the probe and the system, and thus when $I_{\mathrm{E}}$ is determined, we can take $\{\hat{P}_{(k)}\}$ freely.
The QC-mutual information $I_{\mathrm{QC}}$ does not have this freedom.
Theorems 4 and 5 follow from this difference of the freedom between $I_{\mathrm{E}}$ and $I_{\mathrm{QC}}$. 
Thus, in the above context, we can state that the substance of information is the entanglement. 
The information gain is already completed when the unitary interaction is over and the projective measurement is only the interpretation of the information.

\section*{Acknowledgements}
This work was supported by the Grants-in-Aid for Japan Society for Promotion of Science (JSPS) Fellows (Grant No. 24・8116).
The author thanks Prof. Naomichi Hatano for useful discussions.

\appendix

\section{}

In the present appendix, we prove the following theorem:
\begin{theorem}
For an arbitrary three-qubit pure state $\left|\psi_{PSR_{1}}\right\rangle$, there exists a projective measurement $\{\hat{P}_{(k)}\}_{k=0,1}$ such that 
the results $\hat{P}_{(k)}\left|\psi_{PSR_{1}}\right\rangle$ are LU-equivalent for $k=0,1$ and 
\begin{equation}
E^{S\mbox{-}R_{1}}(\hat{P}_{(k)}\left|\psi_{PSR_{1}}\right\rangle)=E^{S\mbox{-}R_{1}}_{F}(\hat{\rho}_{SR_{1}})\label{targetS}
\end{equation}
is valid.
\end{theorem}

\textbf{Proof}:
Because $\hat{\rho}_{SR_{1}}$ is a two-qubit mixed state, we can express $E^{S\mbox{-}R_{1}}_{F}(\hat{\rho}_{SR_{1}})$ in the form of the concurrence \cite{concurrence}:
\begin{equation}
E^{S\mbox{-}R_{1}}_{F}(\hat{\rho}_{SR_{1}})=h\left(\frac{1+\sqrt{1-C^2_{SR_{1}}(\hat{\rho}_{SR_{1}})}}{2}\right),
\end{equation}
where $C_{SR_{1}}(\hat{\rho}_{SR_{1}})$ is the concurrence of $\hat{\rho}_{SR_{1}}$ and $h(x)\equiv-x\mbox{log}x-(1-x)\mbox{log}(1-x)$.
Thus, we only have to find a projective measurement $\{\tilde{P}_{(k)}\}_{k=0,1}$ such that $\tilde{P}_{(k)}\left|\psi_{PSR_{1}}\right\rangle$ for $k=0,1$ are LU-equivalent to each other and $C_{SR_{1}}(\tilde{P}_{(k)}\left|\psi_{PSR_{1}}\right\rangle)=C_{SR_{1}}(\hat{\rho}_{SR_{1}})$.

Before giving the projective measurement $\{\tilde{P}_{(k)}\}$, we first present preparations.
First we express $\left|\psi_{PSR_{1}}\right\rangle$ in the form of the generalized Schmidt decomposition \cite{18}:
\begin{eqnarray}
\left|\psi_{PSR_{1}}\right\rangle=\lambda_{0}\left|000\right\rangle+\lambda_{1}e^{i\varphi}\left|100\right\rangle+\lambda_{2}\left|101\right\rangle\nonumber\\
+\lambda_{3}\left|110\right\rangle+\lambda_{4}\left|111\right\rangle\enskip\enskip
(0\le\varphi\le\pi)\label{L2.1}
\end{eqnarray}
and introduce the following eight parameters \cite{tajima};
\begin{eqnarray}
K_{PS}&\equiv& C^2_{PS}+\tau_{PSR_{1}},\\
K_{PR_{1}}&\equiv& C^2_{PR_{1}}+\tau_{PSR_{1}},\\
K_{SR_{1}}&\equiv& C^2_{SR_{1}}+\tau_{PSR_{1}},\\
\enskip J_{5}&\equiv&4\lambda^2_{0}(|\lambda_{1}\lambda_{4}e^{i\varphi}-\lambda_{2}\lambda_{3}|^2+\lambda^2_{2}\lambda^2_{3}-\lambda^2_{1}\lambda^2_{4}),\\
K_{5}&\equiv& J_{5}+\tau_{PSR_{1}},\\
\Delta_{J}&\equiv& K^2_{5}-K_{PS}K_{PR_{1}}K_{SR_{1}},\\
e^{-i\tilde{\varphi}_{5}}&\equiv&\frac{\lambda_{2}\lambda_{3}-\lambda_{1}\lambda_{4}e^{i\varphi}}{|\lambda_{2}\lambda_{3}-\lambda_{1}\lambda_{4}e^{i\varphi}|},\\
Q_{\mbox{e}}&\equiv&\mbox{sgn}\left[ \sin{\varphi}\left(\lambda^2_{0}-\frac{\tau_{PSR_{1}}+J_{5}}{2(C^2_{SR_{1}}+\tau_{PSR_{1}})}\right)\right],
\end{eqnarray}
where $\tau_{PSR_{1}}$ is the tangle of $\left|\psi_{PSR_{1}}\right\rangle$ and $\mbox{sgn}[x]$ is the sign function,
\begin{eqnarray}
\mbox{sgn}[x] =\left\{ \begin{array}{ll}
x/|x| & (x\ne0) \\
0 & (x=0) \\
\end{array} \right\}.
\end{eqnarray}
When $Q_{\mbox{e}}=0$, there are two possible decompositions which satisfy \eqref{L2.1}.
We then choose the decomposition with a greater coefficient $\lambda_{0}$.

Now we have completed the preparation.
In the basis of \eqref{L2.1}, the projective measurement $\{\tilde{P}_{(k)}\}_{k=0,1}$ is given as follows;
\begin{eqnarray}
\tilde{P}_{(0)}=\sqrt{\left(
\begin{array}{cc}
  a  & ke^{-i\theta} \\ 
   ke^{i\theta} &  b
\end{array}
\right)},\label{surasura5}
\tilde{P}_{(1)}=\sqrt{\left(
\begin{array}{cc}
  1-a  & -ke^{-i\theta} \\ 
   -ke^{i\theta} &  1-b
\end{array}
\right)},\label{surasura6}
\end{eqnarray}
where the measurement parameters $a$, $b$, $k$ and $\theta$ are defined as follows: 
\begin{eqnarray}
a&=&\frac{1}{2}-\frac{K_{5}\tau_{PSR_{1}}\dot\pm\sqrt{\Delta_{J}}C^2_{SR_{1}}}{2K_{SR_{1}}\sqrt{K^2_{5}-K_{PS}K_{PR_{1}}C^2_{SR_{1}}}},\label{a}\\
b&=&1-a,\label{b}\\
k&=&\sqrt{a(1-a)},\label{k}\\
\theta&=&-\tilde{\varphi}_{5},\label{theta}
\end{eqnarray}
and when $Q_{\mbox{e}}\ne0$ the mark $\dot{\pm}$ means $-Q_{\mbox{e}}$ and when $Q_{\mbox{e}}=0$ the mark $\dot{\pm}$ means $-$.  
With using \eqref{a}--\eqref{theta} and Lemma 1 of Ref. \cite{tajima} and after straightforward algebra, we can confirm that the measurement $\{\tilde{P}_{(k)}\}$ is the measurement that we sought.

\renewcommand{\refname}{\vspace{-1cm}}

\end{document}